\documentclass[11pt]{article}
\usepackage{epsfig} 
\newcommand{\sect}[1]{ \section{#1} \setcounter{equation}{0} } 

\newcommand{\half}{\mbox{\small{$\frac{1}{2}$}}}

\newcommand{\Nc}{N_{\!c}}

\newcommand{\pslash}{p \! \! \! /}

\newcommand{\partialslash}{\partial \! \! \! /}
\newcommand{\xslash}{x \! \! \! /}
\newcommand{\yslash}{y \! \! \! /}

\begin{document}
\title{Fermion bilinear operator critical exponents at $O(1/N^2)$ in the 
QED-Gross-Neveu universality class}
\author{J.A. Gracey, \\ Theoretical Physics Division, \\ 
Department of Mathematical Sciences, \\ University of Liverpool, \\ P.O. Box 
147, \\ Liverpool, \\ L69 3BX, \\ United Kingdom.} 
\date{}

\maketitle 

\vspace{5cm} 
\noindent 
{\bf Abstract.} We use the critical point large $N$ formalism to calculate the 
critical exponents corresponding to the fermion mass operator and flavour 
non-singlet fermion bilinear operator in the universality class of Quantum 
Electrodynamics (QED) coupled to the Gross-Neveu model for an $SU(N)$ flavour 
symmetry in $d$-dimensions. The $\epsilon$ expansion of the exponents in 
$d$~$=$~$4$~$-$~$2\epsilon$ dimensions are in agreement with recent three and 
four loop perturbative evaluations of both renormalization group functions of 
these operators. Estimates of the value of the non-singlet operator exponent in
three dimensions are provided. 

\vspace{-17cm}
\hspace{13.0cm}
{\bf LTH 1173}

\newpage 

\sect{Introduction.}

An interesting connection between two different areas of physics has been
developing over recent years. The new material of intense theoretical and
experimental interest in condensed matter physics is that of graphene which is
a one atom thick sheet of Carbon atoms arranged in a hexagonal or honeycomb 
lattice. Under certain conditions, such as when the sheet is deformed by 
stretching, graphene undergoes a transition from a conducting to 
Mott-insulating phase \cite{1,2}. Remarkably it is believed that this phase 
transition is described by the universality class of a Yukawa-type theory which
goes under the general heading of Gross-Neveu or Gross-Neveu-Yukawa theories, 
\cite{3,4,5,6,7}. To understand the properties of such phase transitions 
better, theoretical calculations have been carried out in the various 
Gross-Neveu field theories introduced in \cite{8}. The primary focus in the 
work on several of the classes \cite{9,10,11,12,13,14,15}, for example, was to 
determine estimates for the critical exponents in three dimensions. The main 
methods used in these studies include matched Pad\'{e} approximants based on 
the $\epsilon$-expansion of two and four dimensional field theories in the same 
universality class, the large $N$ method, the functional renormalization group 
technique and Monte Carlo or numerical evaluations. The $\epsilon$-expansion 
approach required computing the underlying renormalization group functions at 
high loop order. For certain specific universality classes there is a general 
consensus on the values of the exponents in three dimensions. See, for 
instance, the comprehensive analysis recently carried out in \cite{15}. In 
other classes such as the chiral Heisenberg Gross-Neveu one the same level of 
precision has yet to be attained. Although there has been some progress 
computationally at four loops, \cite{14}, a matched Pad\'{e} approximant 
estimate in three dimensions would require the same level of precision in the 
two dimensional theory of the universality class. 

While such pure Yukawa type universality classes have been of interest for 
aspects of graphene physics, a second range of classes is also important. These 
are generally defined by adding in a $U(1)$ or Quantum Electrodynamics (QED) 
sector to the pure Gross-Neveu one and are termed the QED-Gross-Neveu 
universality classes. It is this general class which is the bridge between 
condensed matter problems and those in particle physics. This is because the 
underlying quantum field theory is structurally equivalent to that of the 
Standard Model. In its early historical construction the weak interactions were approximated by an effective $4$-fermi operator which was later replaced by a 
gauge-Yukawa class of interactions. Given this there is clearly interest in 
refining our understanding of phase transitions in materials such as graphene 
since, in principle, they could be studied in experimental setups smaller than 
those at CERN, for example, and may increase our knowledge of phase transitions
in the Standard Model or potential theories which lie beyond it. For instance,
there was an indication in \cite{7} that the graphene transition could be a
laboratory for examining the intricacies of the Standard Model spontaneous 
symmetry mechanism. Another example of a recent development concerning phase 
transitions is that of emergent symmetries. In the chiral Ising and chiral XY 
Gross-Neveu universality classes, \cite{16,17,18,19}, there is a fixed point 
where all the critical couplings are equal producing an emergent supersymmetric 
fixed point. The equivalence of critical couplings is a necessary but not 
sufficient condition for this. The additional fact that the anomalous 
dimensions of the bosonic and fermionic fields have the same value supports the
emergent supersymmetry. Potentially there may be extensions of the Standard 
Model or indeed the Standard Model itself where this could also occur. Not all
emergent symmetries correspond to supersymmetry. Instead a theory could be dual
to another which becomes apparent when certain operators have the same critical
exponent at the fixed points in the respective theories. 

One example of such a duality arises in three dimensions between QED and a
critical point version of the $CP^1$ sigma model, \cite{19,20}. More recently 
there has been an extension of this type of duality between the QED-Gross-Neveu
universality class in three dimensions for a specific number of electron
flavours and an $SU(2)$ symmetric non-compact $CP^1$ sigma model. Underlying 
this duality is an emergent $SO(5)$ symmetry at the deconfining quantum 
critical point, \cite{22,23,24}. At present the duality is at the level of a 
conjecture based on numerical evidence and lacks a concrete proof. To study 
this conjecture further two major independent but simultaneous computations 
have been undertaken which involved renormalizing the underlying field theory 
of the QED-Gross-Neveu theory in four dimensions to high loop order. Three loop 
results were determined in \cite{25} with the four loop results following later in \cite{26}. Although the main results of both groups was the construction of 
all the renormalization group functions, one important operator was also 
renormalized which was the flavour non-singlet fermion bilinear which has been 
studied in \cite{27,28}, for instance. Its anomalous dimension critical 
exponent is central to establishing the duality in three dimensions. Therefore 
the $\epsilon$-expansion of the three and four loop exponent at the 
Wilson-Fisher fixed point has to be summed down to three dimensions and an 
estimate determined at $N$~$=$~$1$. Experience from other situations has 
demonstrated that this is not a trivial process. What would be useful in 
contributing to the debate is an independent approach to compute the critical 
exponent of the same operator. That is one aim of this article where we will 
compute the exponent of the operator in the large $N$ expansion at $O(1/N^2)$ 
in $d$-dimensions. The method used the large $N$ critical point approach 
developed in \cite{29,30} for the universality class of $O(N)$ scalar fields 
which contains the two dimensional nonlinear sigma model and four dimensional 
$\phi^4$ theory. The second aim of the article is to determine the flavour
singlet fermion bilinear operator critical exponent at the same order as the
non-singlet one. We will also refer to this as the fermion mass exponent as
this is what the operator corresponds to. It has been given the former term
partly to indicate that there is a large overlap in the computations to
deduce both exponents which will become apparent at $O(1/N^2)$.  

While the critical point large $N$ method has already been applied to the 
QED-Gross-Neveu class at $O(1/N)$, \cite{31,32,33}, it transpires that an early 
$O(1/N^2)$ computation, \cite{32}, of the fermion anomalous dimension exponent,
$\eta$, in the Landau gauge had an error. Therefore, a separate aspect of this 
article is designed to address this failing as the formalism needed for the 
$O(1/N^2)$ operator dimension relies centrally on not only the value of $\eta$ 
but also the $d$-dimensional values of the amplitudes of the propagators in the 
universality class at this order. En route to correct this we will provide an
expression for $\eta$ as a function of the gauge parameter as this is now
important for problems which were not manifest at the time of \cite{32} and 
which we note later. One of the reasons why such an error did not come to light
before was due to a lack of perturbative information to compare with at the 
time of the early work. By this we mean the following. Since the large $N$ 
exponents are determined in the universal theory as a function of the spacetime
dimension $d$, then the coefficients of $\epsilon$ in the power series 
expansion of an exponent at whatever large $N$ order is available when 
$d$~$=$~$4$~$-$~$2\epsilon$ should be in exact agreement with those in the 
perturbative exponent. That was not the case for $\eta$ at $O(1/N^2)$ when 
\cite{25} and \cite{26} became available. So with the recent three and four 
loop results the correct value of $\eta$ can be established. Equally the 
$\epsilon$ expansion of both operator exponents have to be in agreement with 
the corresponding three and four loop critical exponents of \cite{25,26}. 
Therefore our results will partly provide an independent non-trivial check on 
this recent perturbative work. That having been established the next aim can be
addressed which is the restriction of the $d$-dimensional results to three 
dimensions {\em without} a resummation in $\epsilon$. While $O(1/N^2)$ in 
$d$-dimensions represents a level beyond what is normal for usual $1/N$ 
analyses, to estimate a value of the flavour non-singlet operator exponent in 
question for the duality conjecture at $N$~$=$~$1$ may be outside a region of 
applicability. However such an outcome cannot be pre-judged. Moreover, as was 
shown in the analysis of all accumulated analytic knowledge of field theory 
computations in the pure Gross-Neveu universality class, credible exponent 
estimates can emerge for low $N$ by pooling all available data. Therefore our 
$O(1/N^2)$ computation for the flavour non-singlet fermion bilinear operator 
critical exponent should also be viewed in that larger context.

The article is organized as follows. We introduce the large $N$ critical point
formalism for the QED-Gross-Neveu universality class in section $2$. There the
derivation of $\eta$ at $O(1/N^2)$ is given. The next section is devoted to
the $O(1/N^2)$ computation of flavour non-singlet and singlet fermion bilinear 
operator critical exponents in $d$-dimensions. Having established these results
the $\epsilon$ expansions are deduced in section $4$ and compared with the
recent explicit perturbative results. Estimates of the flavour non-singlet 
exponent in three dimensions are also determined for a range of values of $N$ 
in order to compare with the resummation of the perturbative $\epsilon$ 
expansion for the same spacetime. In order to assist such an analysis the
critical exponent for the same operator at $O(1/N^2)$ is also determined and 
studied in the pure Gross-Neveu universality class in section $5$. Concluding 
remarks are provided in section $6$.

\sect{Large $N$ formalism.}

The first step in applying the critical point large $N$ formalism developed in
\cite{29,30} for QED coupled to the Gross-Neveu model is to determine the 
Lagrangian for the underlying universality class at the $d$-dimensional
Wilson-Fisher fixed point. First the pure Gross-Neveu universality class is
the two dimensional quantum field theory given in \cite{8} which corresponds to
an $SU(N)$ multiplet of fermions with a quartic self-interaction. For the 
QED-Gross-Neveu universality class the Lagrangian of \cite{8} is extended to 
include a second quartic self-interaction where this additional term has a 
Thirring model structure. To summarize the two dimensional Lagrangian is given 
by  
\begin{equation}
L^{d=2} ~=~ i \bar{\psi}^i \partialslash \psi^i ~+~
\frac{g_1^2}{2} \left( \bar{\psi}^i \psi^i \right)^2 ~+~
\frac{g_2^2}{2} \left( \bar{\psi}^i \gamma^\mu \psi^i \right)^2
\label{lagd2}
\end{equation}
where $1$~$\leq$~$i$~$\leq$~$N$ and $g_i$ are the two dimensionless coupling
constants in two dimensions. This is not the only way of formulating the 
renormalizable two dimensional theory since two auxiliary fields,
$\tilde{\sigma}$ and $\tilde{A}_\mu$, can be introduced by rewriting 
(\ref{lagd2}) since 
\begin{equation}
L^{d=2} ~=~ i \bar{\psi}^i \partialslash \psi^i ~+~
g_1 \tilde{\sigma} \bar{\psi}^i \psi^i ~+~ 
g_2 \tilde{A}_\mu \bar{\psi}^i \gamma^\mu \psi^i ~-~ 
\frac{1}{2} \tilde{\sigma}^2 ~-~ \frac{1}{2} \tilde{A}_\mu \tilde{A}^\mu ~.
\label{lagd2aux}
\end{equation}
In two dimensions we regard $\tilde{A}_\mu$ as an auxiliary spin-$1$ field 
rather than a photon since there is no gauge symmetry and its propagator is 
unity. Beyond two dimensions this field will become the equivalent of the 
photon at criticality. From (\ref{lagd2aux}) the structure of the underlying 
universal Lagrangian can be deduced. Following the prescription apparent in 
\cite{29,30} the coupling constants of the universal interactions are rescaled 
into the quadratic terms of the auxiliary fields. In this case (\ref{lagd2aux})
becomes  
\begin{equation}
L ~=~ i \bar{\psi}^i \partialslash \psi^i ~+~
\sigma \bar{\psi}^i \psi^i ~+~ A_\mu \bar{\psi}^i \gamma^\mu \psi^i ~-~ 
\frac{1}{2g_1^2} \sigma^2 ~-~ \frac{1}{2g_2^2} A_\mu A^\mu ~.
\label{laguniv}
\end{equation}
The first three terms are core to the Lagrangian of the underlying universal 
theory whereas the remaining two are the only two relevant local operators
when the critical dimension is two. In other critical dimensions other local 
operators will be relevant. To see this a dimensional analysis of
(\ref{laguniv}) implies $\psi$, $\sigma$ and $A_\mu$ have canonical dimensions 
of $\half (d-1)$, $1$ and $1$ respectively in $d$-dimensions. Therefore when 
the critical dimension of the universal theory is four the renormalizable 
Lagrangian which is equivalent to (\ref{lagd2}) and (\ref{lagd2aux}) is
\begin{equation}
L^{d=4} ~=~ i \bar{\psi}^i \partialslash \psi^i ~+~
\frac{1}{2} \partial_\mu \sigma \partial^\mu \sigma ~-~ 
\frac{1}{4} \tilde{F}_{\mu\nu}^2 ~-~
\frac{1}{2b} \left( \partial^\mu \tilde{A}_\mu \right)^2 ~+~ 
\bar{g}_1 \tilde{\sigma} \bar{\psi}^i \psi^i ~+~ 
\bar{g}_2 \tilde{A}_\mu \bar{\psi}^i \gamma^\mu \psi^i ~+~ 
\frac{\bar{g}_3^2}{24} \tilde{\sigma}^4 
\label{lagd4}
\end{equation}
where $\tilde{F}_{\mu\nu}$~$=$~$\partial_\mu \tilde{A}_\nu$~$-$~$\partial_\nu 
\tilde{A}_\mu$, $b$ is the gauge fixing parameter with $b$~$=$~$0$ 
corresponding to the Landau gauge and $\bar{g}_i$ are the three coupling 
constants which are dimensionless in four dimensions. The coupling constants 
have been included with the respective interactions as it is this Lagrangian 
which has recently been renormalized to three and four loops in \cite{25,26} 
respectively. Since the canonical dimensions of $\sigma$ and $A_\mu$ are both 
unity in the universal theory then these fields develop canonical propagators 
in four dimensions. A quartic fermion self-interaction cannot be present since 
that operator would have canonical dimension six in a critical dimension of 
four. Moreover given the structure of the non-gauge sector (\ref{lagd4}) is 
sometimes referred to as the QED-Gross-Neveu-Yukawa model. 

The starting point for computing the $d$-dimensional critical exponents of the
underlying QED-Gross-Neveu universality class in the large $N$ expansion is to
write down the asymptotic scaling forms of the propagators in the approach to
the Wilson-Fisher fixed point. At this point no masses are present and the 
propagators have a power law behaviour. Therefore in $d$-dimensions the
propagators of the three fields of (\ref{laguniv}) have the scaling form, 
\cite{32,33,34,35,36},
\begin{eqnarray}
\langle \psi^i(x) \bar{\psi}^j(y) \rangle & \sim &
\frac{(\xslash-\yslash) A \delta^{ij}}{((x-y)^2)^{\alpha}} \nonumber \\
\langle A_\mu(x) A_\nu(y) \rangle & \sim & 
\frac{B_A}{((x-y)^2)^{\beta_A}} \left[ \eta_{\mu\nu} +
\frac{2( 1 - b )\beta_A}{(2\mu-2\beta_A-1+b)}
\frac{(x-y)_\mu (x-y)_\nu}{(x-y)^2} \right] \nonumber \\
\langle \sigma(x) \sigma(y) \rangle & \sim & 
\frac{B_\sigma}{((x-y)^2)^{\beta_\sigma}} 
\label{propcoord}
\end{eqnarray}
in coordinate space. The quantities $A$, $B_A$ and $B_\sigma$ are coordinate
independent amplitudes and $\alpha$, $\beta_A$ and $\beta_\sigma$ are the full
scaling dimensions of the respective fields. They are defined by
\begin{equation}
\alpha ~=~ \mu ~+~ \half \eta ~~~,~~~
\beta_A ~=~ 1 ~-~ \eta ~-~ \chi_A ~~~,~~~
\beta_\sigma ~=~ 1 ~-~ \eta ~-~ \chi_\sigma 
\end{equation}
where
\begin{equation}
d ~=~ 2 \mu 
\end{equation} 
and $\eta$ is the fermion anomalous dimension. The exponents $\chi_A$ and
$\chi_\sigma$ correspond to the anomalous dimensions of the respective 
$3$-point vertices. Given the nature of the universal theory the anomalous part
of the critical exponent corresponding to the fermion singlet bilinear or mass
operator is related to the anomalous part of $\beta_\sigma$. As we are 
computing directly at the Wilson-Fisher fixed point the critical exponents will
depend only on $\mu$ and $N$ and hence they can be formally expanded in a power
series in $1/N$ such as
\begin{equation}
\eta ~=~ \sum_{n=1}^\infty \frac{\eta_n}{N^n} 
\end{equation} 
and similar notation will also be used for other exponents. The particular 
Lorentz structure of the photon propagators is dictated by the canonical form 
of the propagator in momentum space which is then mapped to this coordinate 
space form by a Fourier transform. As the first stage of the large $N$ method 
to compute exponents is to determine the fermion dimension by solving the 
Schwinger-Dyson equations for the $2$-point functions of the fields in the 
approach to criticality, the asymptotic scaling forms of the $2$-point 
functions are also required. These are deduced by first inverting the momentum 
space forms of the scaling functions which are
\begin{eqnarray}
\langle \psi^i(p) \bar{\psi}^j(-p) \rangle & \sim &
\frac{\pslash \tilde{A} \delta^{ij}}{(p^2)^{\mu-\alpha+1}} 
\nonumber \\
\langle A_\mu(p) A_\nu(-p) \rangle & \sim & 
\frac{\tilde{B}_A}{(p^2)^{\mu-\beta_A}} \left[ \eta_{\mu\nu} -
( 1 - b ) \frac{p_\mu p_\nu}{p^2} \right] \nonumber \\
\langle \sigma(p) \sigma(-p) \rangle & \sim & 
\frac{\tilde{B}_\sigma}{(p^2)^{\mu-\beta_\sigma}} 
\label{propmom}
\end{eqnarray}
where $\tilde{A}$, $\tilde{B}_A$ and $\tilde{B}_\sigma$ are the momentum space
amplitudes. The inverse of these are then mapped back to coordinate space by a
Fourier transform. We therefore have, \cite{34,35,36},
\begin{eqnarray}
\langle \psi^i(x) \bar{\psi}^j(y) \rangle^{-1} & \sim &
\frac{r(\alpha-1) (\xslash-\yslash) \delta^{ij}}{A((x-y)^2)^{2\mu-\alpha+1}} 
\nonumber \\
\langle A_\mu(x) A_\nu(y) \rangle^{-1} & \sim & 
\frac{t(\beta_A)}{B_A((x-y)^2)^{2\mu-\beta_A}} \left[ \eta_{\mu\nu} +
\frac{2(2\mu-\beta_A)}{(2\beta_A-2\mu-1)}
\frac{(x-y)_\mu (x-y)_\nu}{(x-y)^2} \right] \nonumber \\
\langle \sigma(x) \sigma(y) \rangle^{-1} & \sim & 
\frac{p(\beta_\sigma)}{B_\sigma((x-y)^2)^{2\mu-\beta_\sigma}} 
\end{eqnarray}
for our three fields. The process produces different coordinate independent 
amplitudes involving the functions 
\begin{equation}
r(\alpha) ~=~ \frac{\alpha a(\alpha-\mu)}{(\mu-\alpha)a(\alpha)} ~~,~~
p(\alpha) ~=~ \frac{a(\alpha-\mu)}{a(\alpha)} ~~,~~
t(\alpha) ~=~ \frac{[4(\mu-\alpha)^2-1]a(\alpha-\mu)}{4(\mu-\alpha)a(\alpha)} 
\end{equation}
where 
\begin{equation}
a(\alpha) ~=~ \frac{\Gamma(\mu-\alpha)}{\Gamma(\alpha)}
\end{equation}
is used for shorthand, \cite{29}.

{\begin{figure}[hb]
\begin{center}
\includegraphics[width=11cm,height=6.0cm]{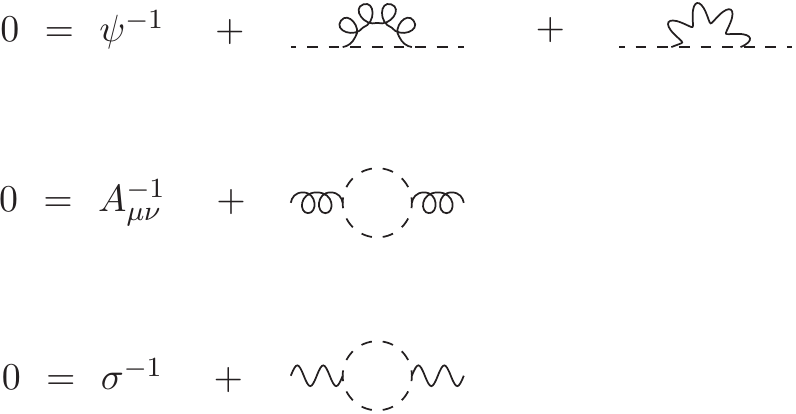}
\end{center}
\caption{Leading order large $N$ graphs contributing to the $2$-point skeleton
Schwinger-Dyson equations.}
\end{figure}}

As we will require basic quantities such as the amplitudes and exponents for
the computation of both bilinear operator exponents at $O(1/N^2)$ and since 
there was an error in the earlier work of \cite{32}, it is instructive to 
review the evaluation of $\eta_2$. The starting point is the set of 
Schwinger-Dyson $2$-point functions in the asymptotic scaling region. The 
leading order contributions in the large $N$ expansion are given in Figure $1$ 
with the higher order corrections to each $2$-point function given in Figures
$2$, $3$ and $4$. It is important to note that the graphs are ordered by powers
of $1/N$ rather than the coupling constant which produces the loop expansion. 
The counting of powers of $N$ is derived by noting that a closed fermion loop 
has a factor of $N$ and each $A_\mu$ and $\sigma$ propagator counts one power 
of $1/N$. Ordinarily in the large $N$ expansion this would produce three loop 
graphs at the same order in $1/N$ as the graphs in Figures $3$ and $4$ where 
there are two closed fermion loops and a total of two $A_\mu$ and $\sigma$ 
propagators. However in (\ref{laguniv}) while such graphs can be present in 
principle we have not included them as either the fermion loop contains an odd 
number of $\gamma$-matrices or graphs with an $A_\mu$ leg vanish by Furry's 
theorem. Therefore we have not included these in Figures $3$ and $4$. For the 
fermion $2$-point function there are also potential three loop graphs with one 
fermion loop but these too are absent for similar reasons. Using 
(\ref{propcoord}) these Schwinger-Dyson equations can be represented 
algebraically by  
\begin{eqnarray}
0 &=& r(\alpha-1) ~+~ z Z_\sigma^2 (x^2)^{\chi_\sigma+\Delta} \nonumber \\
&& -~ \frac{2 y Z_A^2}{(2\mu-3+b)} \left[ [(2\mu-1)(\mu-2)+\mu b]
+ (2\mu-1+b) (1-b) (\eta+\chi_A+\Delta) \right] 
(x^2)^{\chi_A+\Delta} \nonumber \\
&& +~ z^2 \Sigma_1 (x^2)^{2\Delta} ~+~ y^2 \Sigma_2 (x^2)^{2\Delta} ~+~
y z \Sigma_3 (x^2)^{2\Delta} ~+~ y z \Sigma_4 (x^2)^{2\Delta} ~+~
O \left( \frac{1}{N^3} \right) \nonumber \\
0 &=& p(\beta_\sigma) ~+~ 4 N z Z_\sigma^2 (x^2)^{\chi_\sigma+\Delta} ~-~
N z^2 \Gamma_1 (x^2)^{2\Delta} ~-~ N y z \Gamma_2 (x^2)^{2\Delta} ~+~ 
O \left( \frac{1}{N^2} \right) \nonumber \\
0 &=& 2 t(\beta_A) \left[ \frac{(\mu-1)}{(2\mu-1)}
+ \frac{(\eta_1+\chi_{A\,1})}{(2\mu-1)^2N} \right] ~-~ 
4 N y Z_A^2 \left[ \frac{2(\mu-1)}{(2\mu-1)} 
+ \frac{\eta_1}{(2\mu-1)^2N} \right] (x^2)^{\chi_A+\Delta} \nonumber \\
&& -~ N y^2 \left[ \frac{\Pi_1}{\Delta} + \Pi_1^\prime
+ \left[ \frac{\Xi_1}{\Delta} + \frac{\Xi_1^\prime}{2(2\mu-1-\Delta)} \right]
\right] \nonumber \\
&& -~ N y z \left[ \frac{\Pi_2}{\Delta} + \Pi_2^\prime
+ \left[ \frac{\Xi_2}{\Delta} + \frac{\Xi_2^\prime}{2(2\mu-1-\Delta)} \right]
\right] ~+~ O \left( \frac{1}{N^2} \right)
\label{etasd}
\end{eqnarray}
to $O(1/N^2)$. The first two terms of each equation correspond to the three
equations in Figure $1$. An analytic regularization $\Delta$ has been
introduced in (\ref{etasd}) by the shift
\begin{equation}
\chi_A ~ \rightarrow ~ \chi_A ~+~ \Delta ~~~,~~~ 
\chi_\sigma ~ \rightarrow ~ \chi_\sigma ~+~ \Delta 
\end{equation}
since the graphs in Figures $2$, $3$ and $4$ are divergent. Although we are 
working in $d$-dimensional spacetime we are not using dimensional 
regularization. Such a regularization will not quantify the divergences in 
these graphs, \cite{29,30}. Instead this particular analytic regularization is 
used since in this critical point formulation of the large $N$ expansion one is
in effect carrying out perturbation theory in the vertex anomalous dimensions. 
The quantities $\Sigma_i$, $\Gamma_i$, $\Pi_i$ and $\Pi_i^\prime$ represent the
$d$-dependent {\em values} of the graphs devoid of the dimensional dependence 
which has been factored off into the powers of $x^2$ in the correction terms of
(\ref{etasd}). For the photon equation we have formally isolated the divergent 
term in $\Delta$, $\Pi_i$, from the finite part, $\Pi_i^\prime$, of the values 
of both graphs in Figure $3$ because only the transverse part of the 
Schwinger-Dyson equation is relevant in the determination of $\eta$. As the two
$3$-point vertices of (\ref{laguniv}) each involve two fermions and one $A_\mu$
or $\sigma$ field the propagator amplitudes always appear in the combinations 
\begin{equation}
y ~=~ A^2 B_A ~~~,~~~ z ~=~ A^2 B_\sigma ~.
\label{vardef}
\end{equation}
Equally as the underlying $3$-point vertices are always divergent, 
\cite{29,30,37,38}, two vertex renormalization constants $Z_A$ and $Z_\sigma$ 
have been introduced. They only appear on the leading order graphs as the 
effect of the counterterms in the next order will not play a role until 
$\eta_3$ is computed if this method is used.  

\vspace{0.3cm}
{\begin{figure}[hb]
\begin{center}
\includegraphics[width=16.0cm,height=2.0cm]{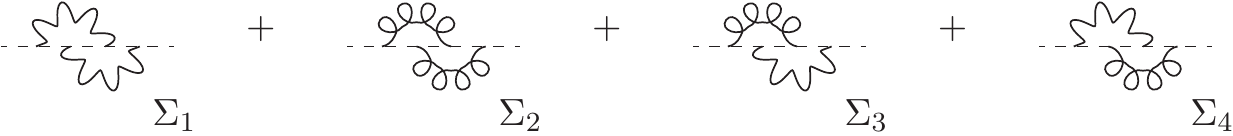}
\end{center}
\caption{$O(1/N^2)$ graphs contributing to fermion $2$-point function.}
\end{figure}}

\vspace{0.3cm}
{\begin{figure}[hb]
\begin{center}
\includegraphics[width=7.5cm,height=1.4cm]{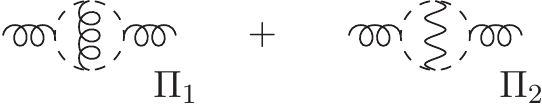}
\end{center}
\caption{$O(1/N^2)$ graphs contributing to the photon $2$-point function.}
\end{figure}}

\vspace{0.3cm}
{\begin{figure}[hb]
\begin{center}
\includegraphics[width=7.5cm,height=1.4cm]{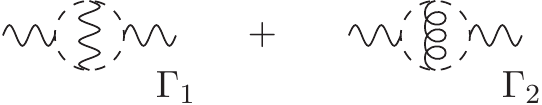}
\end{center}
\caption{$O(1/N^2)$ graphs contributing to the $\sigma$ $2$-point function.}
\end{figure}}

At leading order the three equations simplify to
\begin{eqnarray}
0 &=& r(\alpha-1) ~+~ z ~-~ \frac{2 y}{(2\mu-3+b)} [(2\mu-1)(\mu-2)+\mu b] ~+~ 
O \left( \frac{1}{N^2} \right) \nonumber \\
0 &=& p(\beta_\sigma) ~+~ 4 N z ~+~ O \left( \frac{1}{N^2} \right) \nonumber \\
0 &=& \frac{2 (\mu-1)}{(2\mu-1)} t(\beta_A) ~-~ 
\frac{8(\mu-1)N y}{(2\mu-1)} ~+~ O \left( \frac{1}{N^2} \right)
\label{etasdlo}
\end{eqnarray}
which contain the three unknowns $\eta_1$, $y_1$ and $z_1$. There is no $x^2$
dependence since the expansion of the respective factors in the leading order
terms of (\ref{etasd}) are either $O(1/N)$ or $O(\Delta)$. As there are no
poles in $\Delta$ the regularization can be lifted without any difficulty. This
means that at leading order there is a smooth limit to the critical point as
$x^2$~$\to$~$0$. Setting the canonical values of the exponents for $A_\mu$ and 
$\sigma$ in the final two equations and $\alpha$~$=$~$\mu$~$+$~$\eta_1/N$ in 
the first then the three equations can be solved to produce 
\begin{eqnarray}
\eta_1 &=& -~ [4 \mu^2 - 10 \mu + 5 + (2 \mu - 1) b ] \hat{\eta}_1 \nonumber \\
y_1 &=& 
\frac{(2\mu-1) (2\mu-3+b) \Gamma(2\mu-1)}{16[\mu-1]\Gamma(\mu)\Gamma(1-\mu)} 
\nonumber \\
z_1 &=& -~ \frac{\Gamma(2\mu-1)} {4[\mu-1]\Gamma(\mu)\Gamma(1-\mu)} 
\end{eqnarray}
where 
\begin{equation}
\hat{\eta}_1 ~=~ \frac{\Gamma(2\mu-1)}{4[\mu-1] \Gamma^3(\mu) \Gamma(1-\mu)} ~.
\end{equation}
These agree with the leading order values given in \cite{31,32,33}.

Having established the leading order solution, including the next order terms
is straightforward formally. However there are now five unknowns which are
$\eta_2$, $y_2$, $z_2$, $\chi_{A\,1}$ and $\chi_{\sigma\,1}$. The former three
would be expected given the three leading order equations. The latter two lurk 
within the $1/N$ expansion of the $x$-dependence in the leading order terms. 
Although technically there are two other unknowns which are the vertex 
counterterms their values are not unconnected with the values of $\chi_{A\,1}$ 
and $\chi_{\sigma\,1}$ and determined from the divergent part of the graphs in 
Figures $3$ and $4$. We have computed the explicit values of these graphs to
the finite part in $\Delta$ using the techniques given in 
\cite{29,30,32,33,34,36}. For several graphs the method of conformal 
integration has been applied either directly or by writing the graph as the sum
of scalar integrals after taking traces over fermion lines. The full set of 
values for (\ref{laguniv}) for the $SU(N)$ flavour symmetry and using 
$\mbox{Tr} \, I$~$=$~$4$ for the $\gamma$-matrix trace are
\begin{eqnarray}
\Gamma_1 &=& \frac{8}{(\mu-1)\Gamma^2(\mu)} \left[ \frac{1}{\Delta} ~-~
\frac{1}{(\mu-1)} \right] \nonumber \\
\Gamma_2 &=& \frac{16}{(2\mu-3+b)\Gamma^2(\mu)} 
\left[ \frac{(2\mu-1+b)}{\Delta} + \frac{3}{(\mu-1)}
- 3(\mu-1) \left[ \psi^\prime(\mu-1) - \psi^\prime(1) \right]
\right. \nonumber \\
&& \left. ~~~~~~~~~~~~~~~~~~~~~~~~~
- \frac{2(2\mu-1+b)}{(2\mu-3+b)} \right] \nonumber \\
\Sigma_1 &=& -~ \frac{2}{(\mu-1)\Gamma^2(\mu)} \left[ \frac{1}{\Delta} ~-~
\frac{1}{2(\mu-1)} \right] \nonumber \\
\Sigma_2 &=& \frac{2}{\mu(2\mu-3+b)^2\Gamma^2(\mu)} 
\left[ \frac{[(2\mu-1)(\mu-2) + \mu b]^2}{\Delta} 
- \frac{8 [ (2 \mu-1) (\mu-2) + \mu b ]^2}{[2 \mu-3+b]}
\right. \nonumber \\
&& \left. ~~~~~~~~~~~~~~~~~~~~~~~~~~~~
+~ 4 (\mu-1)^2 \left[ \frac{2 (\mu-1)^2}{\mu} + \frac{1}{2} (2 \mu-3)
+ \frac{(\mu+1)b}{2[\mu-1]} \right]
\right. \nonumber \\
&& \left. ~~~~~~~~~~~~~~~~~~~~~~~~~~~~
-~ 4 (1-b) (2 \mu-1-\mu b)
- (1-b) [ (2 \mu-1) (\mu-2) + \mu b ] \right] \nonumber \\
\Sigma_3 &=& \Sigma_4 ~=~ \frac{1}{\Gamma^2(\mu)} 
\left[ -~ \frac{4[(2\mu-1)(\mu-2) + \mu b]}{\mu(2\mu-3+b)\Delta} ~+~ 
\frac{2(\mu^2+\mu-1)}{\mu^2(\mu-1)}
\right. \nonumber \\
&& \left. ~~~~~~~~~~~~~~~~~~~
+~ \frac{2(2\mu-1)(1-b)}{\mu^2(\mu-1)(2\mu-3+b)} 
~+~ \frac{4(2\mu-1)(1-b)}{\mu(\mu-1)(2\mu-3+b)^2} \right] \nonumber \\
\Xi_i &=& -~ 2 \Pi_i ~~~,~~~ \Xi_i^\prime ~=~ -~ 2 \Pi_i^\prime \nonumber \\
\Pi_1 &=& -~ \frac{16 [(2\mu-1)(\mu-2)+\mu b]}{\mu(2\mu-3+b)\Gamma^2(\mu)}
\nonumber \\
\Pi_1^\prime &=& \frac{16}{(2\mu-3+b)\Gamma^2(\mu)}
\left[ 3(\mu-1) \left[ \psi^\prime(\mu-1) - \psi^\prime(1) \right] ~-~
\frac{3}{(\mu-1)}
\right. \nonumber \\
&& \left. ~~~~~~~~~~~~~~~~~~~~~~~~
~+~ \frac{2[(2\mu-1)(\mu-2) + \mu b]}{\mu(2\mu-3+b)} \right] \nonumber \\
\Pi_2 &=& \frac{8}{\mu\Gamma^2(\mu)} ~~~~,~~~~ 
\Pi_2^\prime ~=~ -~ \frac{8}{\mu(\mu-1)\Gamma^2(\mu)}
\end{eqnarray}
where $\psi(z)$ is the Euler polygamma function. With these the vertex
counterterms and values for $\chi_A$ and $\chi_\sigma$ can be deduced. The
former are chosen in a minimal subtraction scheme by ensuring that there are
no poles in $\Delta$ in the respective $A_\mu$ and $\sigma$ $2$-point
functions. A check on the resultant values 
\begin{eqnarray}
Z_A &=& 1 ~-~ \frac{\eta_1}{2\Delta N} ~+~ O \left( \frac{1}{N^2} \right)
\nonumber \\
Z_\sigma &=& 1 ~-~ 
[ 4 \mu^2 - 6 \mu + 3 + (2\mu-1) b ] \frac{\hat{\eta}_1}{2 \Delta N} ~+~ 
O \left( \frac{1}{N^2} \right)
\end{eqnarray}
is effected by noting that these choices also render the fermion $2$-point
function finite where both renormalization constants are present. Otherwise if
this consistency check was not satisfied we would be working with a
non-renormalizable critical Lagrangian. The values of $\chi_A$ and
$\chi_\sigma$ are determined by ensuring that there are no $\ln(x^2)$ terms in
the now $\Delta$-finite Schwinger-Dyson equations. Again the respective 
exponents are deduced from the $A_\mu$ and $\sigma$ equations with the fermion 
equation used as a check. Consequently  
\begin{equation}
\chi_{\sigma \,1} ~=~ -~ \left[ 4 \mu^2 - 6 \mu + 3 + (2\mu-1) b \right]
\hat{\eta}_1 ~~~,~~~
\chi_{A\,1} ~=~ -~ \eta_1
\end{equation}
which are not unrelated to the renormalization constants. The second relation
is the manifestation of the Ward-Takahashi identity in the critical point
formalism for (\ref{laguniv}) similar to the situation in the pure QED case, 
\cite{31,32,35}. It implies that $A_\mu$ has a dimension of $1$ for all values 
of $N$ in the universal theory. We did not assume this at the outset as its 
emergence acts as another internal consistency check.  

Once these initial leading order quantities are determined the Schwinger-Dyson
equations are both finite and the $x^2$~$\to$~$0$ limit can be smoothly taken
to leave the three equations analogous to (\ref{etasdlo}) from which the three
remaining variables can be determined. We find 
\begin{eqnarray}
y_2 &=& -~ \left[
\left[ ( 4 \mu^3 - 10 \mu^2 + \mu + 2 ) (2 \mu-3+b)
- 4 (2 \mu-1) (\mu-1)^2 \right] \frac{(2 \mu-1)}{4\mu[\mu-1]}
\right. \nonumber \\
&& \left. ~~~~
+ \frac{3}{4} (2 \mu-3+b) (2 \mu-1)^2 (\mu-1) 
\left[ \psi^\prime(\mu-1) - \psi^\prime(1) \right]
\right] \Gamma^2(\mu) \hat{\eta}_1^2 \nonumber \\
z_2 &=& \left[ [ 4 \mu^2 - 14 \mu + 7 ] 
+ \frac{4 (2 \mu-1) (\mu-1) (\mu-1)}{[2 \mu-3+b]}
\right. \nonumber \\
&& \left. ~
-~ 2 (2 \mu-1) (\mu-1)^2
\left[  \psi(2 \mu-1) - \psi(1) + \psi(1-\mu) - \psi(\mu-1) \right]
\right. \nonumber \\
&& \left. ~
+~ 3 (2 \mu-1) (\mu-1)^3 \left[ \psi^\prime(\mu-1) - \psi^\prime(1) \right]
\right] \Gamma^2(\mu) \hat{\eta}_1^2
\end{eqnarray}
for the two amplitude combinations leaving 
\begin{eqnarray}
\eta_2 &=& \left[
\frac{[8 \mu^5 - 60 \mu^4 + 136 \mu^3 - 123 \mu^2 + 50 \mu - 8](2 \mu-1)}
{\mu^2 [\mu-1]} 
+ \frac{[4 \mu^2 - 11 \mu + 4] (2 \mu-1)^2 b}{\mu [\mu-1]} 
\right. \nonumber \\
&& \left. ~ -~ \frac{4 (2 \mu-1) (\mu-1)^2}{\mu}
\left[ \psi(2 \mu-1) - \psi(1) + \psi(1-\mu) - \psi(\mu-1) \right]
\right. \nonumber \\
&& \left. ~ +~ 3 (2 \mu-1) (\mu-1) [4 \mu^2-10 \mu+5+(2 \mu-1) b]
\left[ \psi^\prime(\mu-1) - \psi^\prime(1) \right]
\right] \hat{\eta}_1^2 ~. 
\label{eta2}
\end{eqnarray}
As a check on this expression and that for $\eta_1$ we have computed the 
$\epsilon$ expansion in $d$~$=$~$4$~$-$~$2\epsilon$ dimensions to produce 
\begin{eqnarray}
\left. \eta \right|_{d=4-2\epsilon} &=& 
\left[
\frac{1}{2} \epsilon
- 3 \epsilon^2
+ \frac{3}{2} \epsilon^3
+ \left[ 2 + \zeta_3 \right] \epsilon^4
+ \left[ \frac{3}{2} \zeta_4 
- 6 \zeta_3
+ \frac{5}{2} 
\right] \epsilon^5
+ \left[ 3 \zeta_5
+ 3 \zeta_3
- 9 \zeta_4
+ 3 \right] \epsilon^6
\right] \! \frac{1}{N}
\nonumber \\
&&
+ \left[
\frac{3}{2} \epsilon
- 10 \epsilon^2
+ \left[ \frac{117}{8}
+ \frac{9}{2} \zeta_3
\right] \epsilon^3
+ \left[ \frac{169}{16}
- \frac{45}{2} \zeta_3
+ \frac{27}{2} \zeta_4
\right] \epsilon^4
\right. \nonumber \\
&& \left. ~~~~
- \left[ \frac{261}{32}
- 4 \zeta_3
+ \frac{135}{4} \zeta_4
- 9 \zeta_5 \right] \epsilon^5
\right. \nonumber \\
&& \left. ~~~~
- \left[ \frac{1535}{64}
- 18 \zeta_3^2
- 57 \zeta_3
- 6 \zeta_4
+ 27 \zeta_5
- \frac{45}{4} \zeta_6 \right] \epsilon^6
\right] \frac{1}{N^2} ~+~ O \left( \epsilon^7, \frac{1}{N^3} \right)
\end{eqnarray}
in the Landau gauge where $\zeta_z$ is the Riemann zeta function. We recall 
that the Landau gauge is the fixed point of the renormalization group flow of 
the gauge parameter which can be regarded as a coupling. Comparing with the 
recent three and four loop expressions given in \cite{25,26} we find exact 
agreement. Therefore (\ref{eta2}) now supersedes the $b$~$=$~$0$ result of 
\cite{32}. Consequently we note that in three dimensions
\begin{equation}
\left. \eta \right|_{d=3} ~=~ ~-~ \frac{2 [ 1 - 2 b ]}{\pi^2 N} ~-~ 
\frac{[ ( 36 - 72 b ) \pi^2 + ( 672 b - 256 ) ]}{9 \pi^4 N^2} ~+~ 
O \left( \frac{1}{N^3} \right)
\end{equation}
or 
\begin{equation}
\left. \eta \right|_{d=3} ~=~ -~
\frac{[ 0.202642 - 0.405285 b ]}{N} ~+~ 
\frac{[ 0.044043 b - 0.113275 ]}{N^2} ~+~ O \left( \frac{1}{N^3} \right)
\end{equation}
numerically. The gauge parameter is included here as it is relevant for studies
of chiral symmetry breaking in three dimensions using the large $N$ expansion,
\cite{39}. See, for instance, recent $O(1/N^2)$ work in this area in pure QED, 
\cite{40,41,42}. There the value of $N$ corresponding to chiral symmetry
breaking was studied using $1/N$ methods and the value shown to be independent 
of the gauge. 

A final part of the exercise in reconstructing $\eta_2$ is to lay the 
foundation for determining the $O(1/N^2)$ exponents of the fermion bilinear 
operators. As this will be computed using the critical propagators in momentum 
space, (\ref{propmom}), we need to record the corresponding momentum space 
variables to $O(1/N^2)$ which are defined in a similar way to (\ref{vardef}) by 
\begin{equation}
\tilde{y} ~=~ \tilde{A}^2 \tilde{B}_A ~~~,~~~
\tilde{z} ~=~ \tilde{A}^2 \tilde{B}_\sigma ~.
\end{equation}
We found 
\begin{eqnarray}
\tilde{y}_1 &=& -~ \frac{(2\mu-1) \Gamma(2\mu-1)}
{8(\mu-1) \Gamma^2(\mu) \Gamma(1-\mu)} \nonumber \\
\tilde{z}_1 &=& \frac{\Gamma(2\mu-1)} {4 \Gamma^2(\mu) \Gamma(1-\mu)} 
\end{eqnarray}
at leading order and 
\begin{eqnarray}
\tilde{y}_2 &=& \frac{3}{2} \left[ (2 \mu-1)^2 (\mu-1) 
\left[ \psi^\prime(\mu-1) - \psi^\prime(1) \right]
- \frac{(2\mu-1)^2}{[\mu-1]} \right] \Gamma(\mu) \hat{\eta}_1^2
\nonumber \\
\tilde{z}_2 &=& \left[ (2 \mu-1)
+ 2 (2 \mu-1) (\mu-1)
\left[ \psi(2 \mu-1) - \psi(1) + \psi(1-\mu) - \psi(\mu-1) \right]
\right. \nonumber \\
&& \left. ~
-~ 3 (2 \mu-1) (\mu-1)^2 \left[ \psi^\prime(\mu-1) - \psi^\prime(1) \right]
\right] \Gamma(\mu) \hat{\eta}_1^2
\label{momamp}
\end{eqnarray}
at next order. These were deduced from the momentum space version of 
(\ref{etasd}). 

\sect{Operator critical exponents.}

We now turn to the evaluation of the operator critical exponents which are
those for the gauge invarianr flavour non-singlet and singlet fermion bilinears
\begin{equation}
{\cal O}_m ~=~ \bar{\psi}^i \psi^i ~~~,~~~ 
{\cal O}_{\mbox{ns}} ~=~ \bar{\psi}^i \sigma^z_{ij} \psi^j
\label{opdef}
\end{equation}
where we use the notation of \cite{25,28} for the latter. In \cite{26} the 
notation used for the non-singlet operator was $\bar{\psi}^i T^a_{ij} \psi^j$. 
In each case $\sigma^z$ and $T^a$ are flavour matrices and their presence is to
distinguish the operator from the usual mass operator $\bar{\psi}^i \psi^i$ 
which is the flavour singlet partner quantity. In terms of the full critical 
exponent of both operators they are each comprised of two parts and defined by 
\begin{equation}
\Delta_m ~=~ 2 \mu ~-~ 1 ~+~ \eta ~+~ \eta_{{\cal O}_m} ~~~,~~~ 
\Delta_{\bar{\psi}\sigma^z\psi} ~=~ 2 \mu ~-~ 1 ~+~ \eta ~+~ 
\eta_{{\cal O}_{\mbox{ns}}}
\label{opexpdef}
\end{equation}
where $\eta_{{\cal O}_{\mbox{ns}}}$ is determined to $O(1/N^2)$ by computing 
the leading and next to leading order set of graphs given in Figures $5$ and 
$6$. For the flavour non-singlet operator there are no $O(1/N^2)$ graphs where 
the operator is inserted in a closed fermion loop as a trace over the flavour 
indices would produce either $\mbox{Tr} \, \sigma^z$ or $\mbox{Tr} \, T^a$ 
which vanish. For the mass operator such graphs would have to be included and 
these are given in Figure $7$. As an aside the comparison of the mass operator 
dimension at criticality with the perturbative fermion mass anomalous dimension
at criticality is not straightforward. This is because in the four dimensional 
Lagrangian the mass operator has the same canonical dimension, which is $3$, as
the operator is $\sigma^3$. Therefore under renormalization in the coupling 
constant expansion there is mixing. So in order to compare with the large $N$ 
exponents one has to first compute the anomalous dimensions of the two 
eigen-operators of the perturbative mixing matrix. This was carried out in 
\cite{26}. Evaluating these at the fixed point, one of the eigen-exponents will
correspond to the fermion mass critical exponent computed in the large $N$ 
expansion. By contrast when one computes using the large $N$ critical point 
formalism directly at the Wilson-Fisher fixed point the fermion mass operator 
and $\sigma^3$ have {\em different} canonical dimensions which are $(2\mu-1)$ 
and $3$ respectively. Therefore there is no mixing in the $1/N$ approach. The
canonical dimensions only agree in four dimensions corresponding to the 
operator mixing of perturbation theory. We have mentioned this subtlety at 
length as there may be a potential mixing in the non-singlet situation. 
However, given the nature of the operator (\ref{opdef}) which is a flavour 
vector there is no corresponding flavour non-singlet operator involving three 
$\sigma$ fields. While we will determine the mass operator dimension given the 
nature of the underlying Lagrangian the computation is equivalent to finding 
the $O(1/N^2)$ corrections to $\chi$. This was the method used to deduce the 
fermion mass in the pure Gross-Neveu model, \cite{43}. 

\vspace{0.3cm}
{\begin{figure}[ht]
\begin{center}
\includegraphics[width=7cm,height=2.5cm]{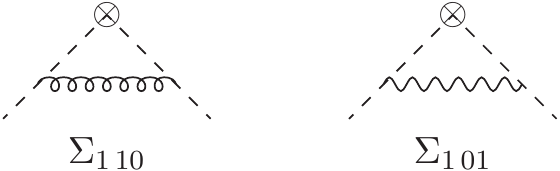}
\end{center}
\caption{Leading order graphs contributing to the critical exponent of
${\cal O}_{\mbox{ns}}$.}
\end{figure}}

First we focus on the determination of $\eta_{{\cal O}_{\mbox{ns}}}$ at 
$O(1/N^2)$. We have computed the graphs of Figures $5$ and $6$ in momentum 
space using the critical propagators (\ref{propmom}) which required the new 
amplitudes (\ref{momamp}) of the previous section. Throughout this section all 
the results will be solely in the Landau gauge as this gauge is a fixed point 
of the large $N$ critical point formalism we are using. While we work in 
momentum space the procedure follows that outlined for determining $\eta$ 
except that the vertex counterterms are already available when the $O(1/N^2)$
corrections to the leading graphs of Figure $5$ are computed. Equally the 
unknown exponent $\eta_{{\cal O}_{\mbox{ns}}i}$ is deduced at each order from 
ensuring that the $p^2$~$\rightarrow$~$0$ limit can be smoothly taken which 
means that there has to be no $\ln p^2$ terms. Finally as there is no mixing 
and there are no derivatives in the operator itself the graphs of Figures $5$ 
and $6$ can be calculated where the operator is inserted at zero momentum. This
means that in effect all the graphs reduce to $2$-point ones so that to 
evaluate the corrections the same conformal integration methods are used as 
those for the graphs giving $\eta_2$. At leading order the two one loop graphs 
are straightforward to evaluate and lead to
\begin{equation}
\eta_{{\cal O}_{\mbox{ns}}1} ~=~ [4 \mu^2 - 6 \mu + 3 ] \hat{\eta}_1
\end{equation}
or 
\begin{equation}
\eta_1 ~+~ \eta_{{\cal O}_{\mbox{ns}}1} ~=~ 2 [2 \mu - 1 ] \hat{\eta}_1
\end{equation}
for (\ref{opexpdef}). 

To complete the $O(1/N^2)$ computation for the non-singlet operator dimension 
evaluation we note that the contributions from the higher order graphs are  
\begin{eqnarray}
\Sigma_{2\,20} &=& -~ \frac{4(2\mu^3-6\mu^2+4\mu-1)(2\mu-1)}
{\mu^2(\mu-1)\Gamma^2(\mu)} ~~~,~~~ 
\Sigma_{2\,11a} ~=~ \frac{2(2\mu-1)^2}{\mu^2\Gamma^2(\mu)}
\nonumber \\
\Sigma_{2\,11b} &=& -~ \frac{4(2\mu-1)}{(\mu-1)\Gamma^2(\mu)} ~~~,~~~ 
\Sigma_{2\,02} ~=~ -~ \frac{4}{(\mu-1)\Gamma^2(\mu)}
\nonumber \\
\Sigma_{4\,20} &=& -~ \frac{2(2\mu-1)^2}{3(\mu-1)\Gamma^2(\mu)} ~~~,~~~ 
\Sigma_{4\,11a} ~=~ \Sigma_{4\,11b} ~=~
-~ \frac{2(2\mu-1)}{3(\mu-1)\Gamma^2(\mu)}
\nonumber \\
\Sigma_{4\,02} &=& -~ \frac{2}{3(\mu-1)\Gamma^2(\mu)} ~~~,~~~
\Sigma_{5\,20} ~=~ -~ \frac{2(2\mu-1)(2\mu-3)}{(\mu-1)\Gamma^2(\mu)}
\nonumber \\
\Sigma_{5\,11} &=& -~ \frac{2(2\mu-1)}{(\mu-1)\Gamma^2(\mu)} ~~~,~~~ 
\Sigma_{5\,02} ~=~ \frac{2}{(\mu-1)\Gamma^2(\mu)}
\label{opval}
\end{eqnarray}
in the Landau gauge. While several of these graphs have been determined in 
either the pure QED or Gross-Neveu cases we have included them all here for
completeness and also since they have been recalculated using with the same 
spinor trace conventions. Different conventions were used in the original 
separate cases. Equally we have written a routine in the symbolic manipulation 
language {\sc Form} \cite{44,45} to evaluate the Feynman integrals. Each of the
graphs is decomposed into a sum of scalar integrals after taking the trace and 
then computed separately. While several of these scalar integrals were already 
known for the pure QED and Gross-Neveu cases new ones had to be calculated such
as the set $\Sigma_{i\, 11}$. Equipped with (\ref{opval}) we find 
\begin{eqnarray}
\eta_{{\cal O}_{\mbox{ns}}2} &=& -~ \left[
\frac{[ 8 \mu^4 - 44 \mu^3 + 56 \mu^2 - 27 \mu + 4 ] (2 \mu-1)}{\mu [\mu-1]}
\right. \nonumber \\
&& \left. ~~~~~ 
+~ 4 (2 \mu-1) (\mu-1)
\left[ \psi(2 \mu-1) - \psi(1) + \psi(1-\mu) - \psi(\mu-1) \right]
\right. \nonumber \\
&& \left. ~~~~~
+~ 3 [ 4 \mu^2 - 6 \mu + 3 ] (2 \mu-1) (\mu-1)
\left[ \psi^\prime(\mu-1) - \psi^\prime(1) \right]
\right] \hat{\eta}_1^2
\end{eqnarray}
which produces 
\begin{eqnarray}
\eta_2 ~+~ \eta_{{\cal O}_{\mbox{ns}}2} &=& -~ \left[  
\frac{2 [ 4 \mu^3 - 18 \mu^2 + 15 \mu - 4 ] (2 \mu-1)}{\mu^2 [\mu-1]}
\right. \nonumber \\
&& \left. ~~~~~ 
+~ \frac{4 (2 \mu-1)^2 (\mu-1)}{\mu}
\left[ \psi(2 \mu-1) - \psi(1) + \psi(1-\mu) - \psi(\mu-1) \right]
\right. \nonumber \\
&& \left. ~~~~~
+~ 6 (2 \mu-1)^2 (\mu-1) \left[ \psi^\prime(\mu-1) - \psi^\prime(1) \right]
\right] \hat{\eta}_1^2
\label{opdim}
\end{eqnarray}
which is one of the main results of this article and we recall that 
$d$~$=$~$2\mu$ and $\psi(z)$ is the polygamma function.

\vspace{0.3cm}
{\begin{figure}[ht]
\begin{center}
\includegraphics[width=15.0cm,height=15.0cm]{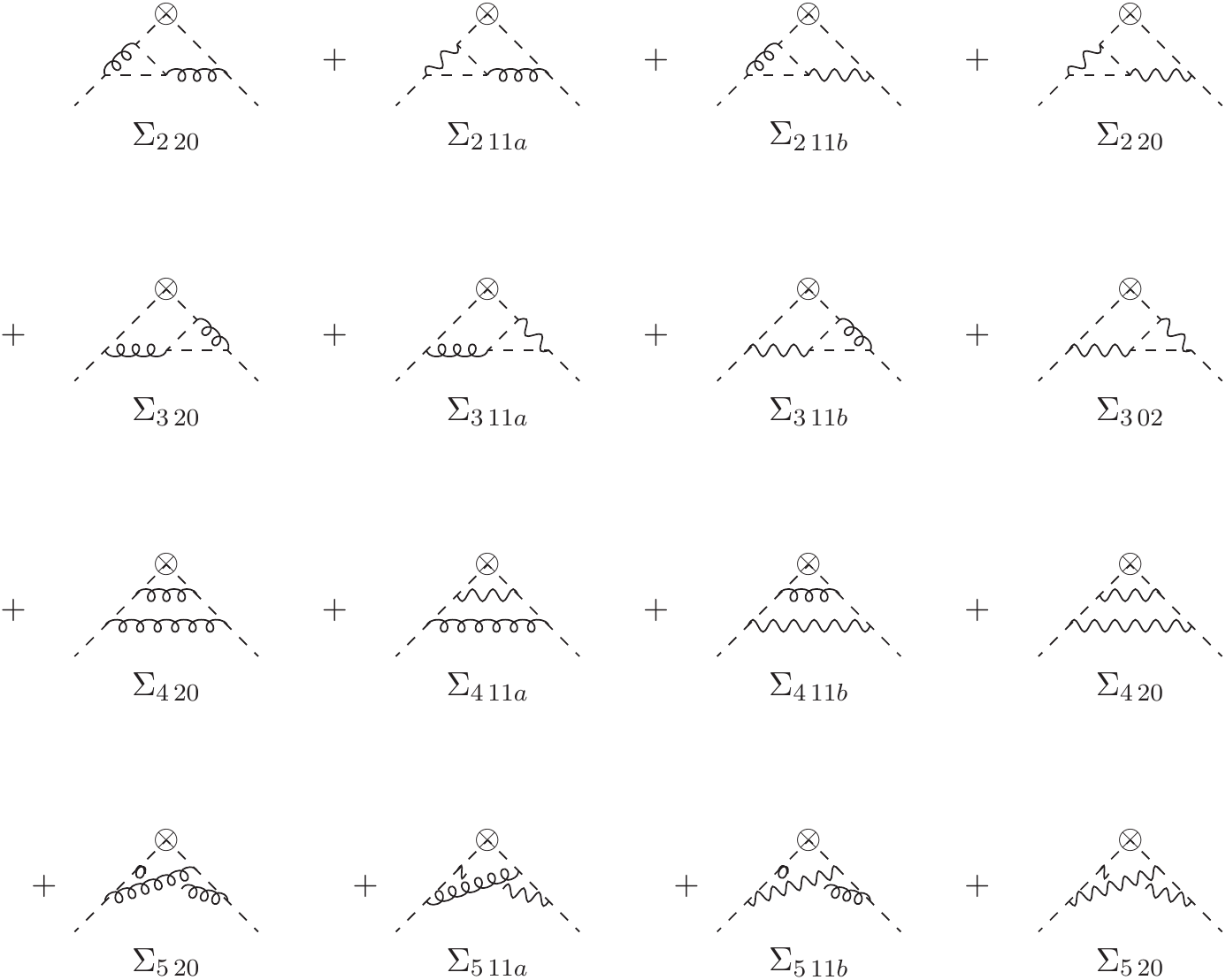}
\end{center}
\caption{$O(1/N^2)$ graphs contributing to the critical exponent of
${\cal O}_{\mbox{ns}}$.}
\end{figure}}

The situation for the mass operator is similar to that of the flavour
non-singlet case except that the values of the additional graphs of Figure
$7$ have to be included. Using similar conformal integration techniques as 
those used in \cite{43} produces 
\begin{eqnarray}
\Sigma_{6\,000} &=& -~ \frac{4[2\mu^2-5\mu+4]\Gamma(1-\mu)}
{3(\mu-1)\Gamma(2\mu-1)} 
\nonumber \\
\Sigma_{6\,110} &=& \Sigma_{6\,011} ~=~ 
-~ \left[ \frac{2[8\mu^3-28\mu^2+34\mu-17]}{3(\mu-1)}
+ 4(\mu-1) \left[ \psi^\prime(\mu-1) - \psi^\prime(1) \right] \right] 
\frac{\Gamma(1-\mu)}{\Gamma(2\mu-1)} 
\nonumber \\
\Sigma_{6\,011} &=& -~ \frac{[8\mu^2-3\mu+3]\Gamma(1-\mu)} 
{3(\mu-1)\Gamma(2\mu-1)} 
\nonumber \\
\Sigma_{7\,000} &=& \left[ 
4(\mu-1) \left[ \psi^\prime(\mu-1) - \psi^\prime(1) \right] 
- \frac{4\mu}{3(\mu-1)} \right] 
\frac{\Gamma(1-\mu)}{\Gamma(2\mu-1)} 
\nonumber \\
\Sigma_{7\,110} &=& \Sigma_{7\,101} ~=~ 
\left[ \frac{4[2\mu^2-2\mu-1]}{3(\mu-1)}
- 4(\mu-1)^2 \left[ \psi^\prime(\mu-1) - \psi^\prime(1) \right] \right] 
\frac{\Gamma(1-\mu)}{\Gamma(2\mu-1)} 
\nonumber \\
\Sigma_{7\,011} &=& \left[
4(\mu-1)(\mu-2) \left[ \psi^\prime(\mu-1) - \psi^\prime(1) \right] 
- \frac{2[4\mu^2-2\mu-9]}{3(\mu-1)} \right] 
\frac{\Gamma(1-\mu)}{\Gamma(2\mu-1)} 
\end{eqnarray}
in the Landau gauge. At leading order in $1/N$ since the set of graphs is the
same for the extraction of the non-singlet operator exponent then 
\begin{equation}
\eta_{{\cal O}_{m \, 1}} ~=~ \eta_{{\cal O}_{\mbox{ns}}1} ~.
\end{equation}
A similar relation does not hold at next order due to the graphs of Figure $7$
but these lead to 
\begin{eqnarray}
\eta_{{\cal O}_m \, 2} &=& -~ \left[
\frac{[ 48 \mu^6 - 208 \mu^5 + 304 \mu^4 - 186 \mu^3 + 16 \mu^2 + 23 \mu - 4 ]}
{\mu[\mu-1]}
\right. \nonumber \\
&& \left. ~~~~~ 
+~ 4 (2 \mu-1) (\mu-1)
\left[ \psi(2 \mu-1) - \psi(1) + \psi(1-\mu) - \psi(\mu-1) \right]
\right. \nonumber \\
&& \left. ~~~~~
+~ 3 [ 16 \mu^3 - 20 \mu^2 + 14 \mu - 5 ] (\mu-1)
\left[ \psi^\prime(\mu-1) - \psi^\prime(1) \right]
\right] \hat{\eta}_1^2
\end{eqnarray}
which implies 
\begin{eqnarray}
\eta_2 ~+~ \eta_{{\cal O}_m \, 2} &=& -~ \left[  
\frac{2 [ 24 \mu^7 - 112 \mu^6 + 216 \mu^5 - 259 \mu^4 + 199 \mu^3 - 100 \mu^2
+ 31 \mu - 4 ]}{\mu^2[\mu-1]}
\right. \nonumber \\
&& \left. ~~~~~ 
+~ \frac{4 (2 \mu-1)^2 (\mu-1)}{\mu}
\left[ \psi(2 \mu-1) - \psi(1) + \psi(1-\mu) - \psi(\mu-1) \right]
\right. \nonumber \\
&& \left. ~~~~~
+~ 6 [ 4\mu^2 + 2 \mu - 3] \mu (\mu-1) 
\left[ \psi^\prime(\mu-1) - \psi^\prime(1) \right]
\right] \hat{\eta}_1^2 
\label{massdim}
\end{eqnarray}
where we note that $\chi_{\sigma \, i}$~$=$~$\eta_{{\cal O}_m \, i}$.

\vspace{0.3cm}
{\begin{figure}[ht]
\begin{center}
\includegraphics[width=15.0cm,height=7.0cm]{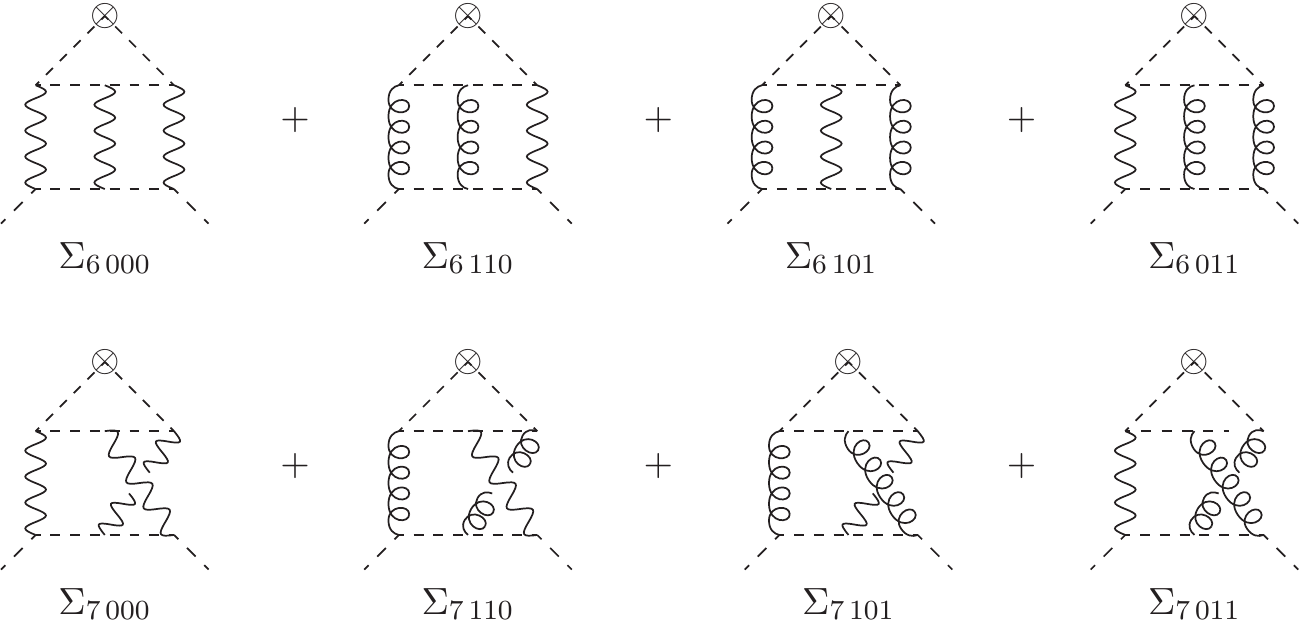}
\end{center}
\caption{Additional $O(1/N^2)$ graphs for mass operator critical exponent.} 
\end{figure}}

\sect{Results.}

This section is devoted to an analysis of the exponents at $O(1/N^2)$. As a 
first stage we note that the $\epsilon$ expansion of each of (\ref{opdim}) and 
(\ref{massdim}) near four dimensions is in total agreement with the recent 
three and four loop perturbative computations of \cite{25,26}. More explicitly 
we note 
\begin{eqnarray}
\left. \Delta_m \right|_{d=4-2\epsilon} &=&
3 ~-~ 2 \epsilon
\nonumber \\
&&
+ \left[
-~ 3 \epsilon
+ 2 \epsilon^2
+ 3 \epsilon^3
+ \left[ 4  
- 6 \zeta_3 \right] \epsilon^4
+ \left[ 5
- 9 \zeta_4 
+ 4 \zeta_3 \right] \epsilon^5
\right. \nonumber \\
&& \left. ~~~~
+ \left[ 6 
- 18 \zeta_5  
+ 6 \zeta_4 
+ 6 \zeta_3 \right] \epsilon^6
\right]
\frac{1}{N} 
\nonumber \\
&&
+~ \left[ 
\frac{9}{2} \epsilon
- \frac{39}{4} \epsilon^2
+ \left[ \frac{413}{8}
- 102 \zeta_3 \right] \epsilon^3
+ \left[ -~ \frac{1413}{16}
- 153 \zeta_4
+ 297 \zeta_3 \right] \epsilon^4
\right. \nonumber \\
&& \left. ~~~~
+ \left[ -~ \frac{1935}{32}
- 204 \zeta_5
+ \frac{891}{2} \zeta_4 
- 105 \zeta_3 \right] \epsilon^5
\right. \nonumber \\
&& \left. ~~~~
+ \left[ -~ \frac{4017}{64}
- 255 \zeta_6 
+ 648 \zeta_5 
- \frac{315}{2} \zeta_4 
+ 162 \zeta_3 
- 408 \zeta_3^2 \right] \epsilon^6
\right]   
\frac{1}{N^2}
\nonumber \\
&& +~ O \left( \epsilon^7, \frac{1}{N^3} \right)
\end{eqnarray}
for the mass operator where $\zeta_z$ is the Riemann zeta function and 
\begin{eqnarray}
\left. \Delta_{\bar{\psi}\sigma^z\psi} \right|_{d=4-2\epsilon} &=&
3 ~-~ 2 \epsilon
\nonumber \\
&&
+ \left[
-~ 3 \epsilon
+ 2 \epsilon^2
+ 3 \epsilon^3
+ \left[ 4
- 6 \zeta_3 \right] \epsilon^4
+ \left[ 5
+ 4 \zeta_3
- 9 \zeta_4 \right] \epsilon^5
\right. \nonumber \\
&& \left. ~~~~
+ \left[ 6
+ 6 \zeta_3
+ 6 \zeta_4
- 18 \zeta_5 \right] \epsilon^6
\right] \frac{1}{N}
\nonumber \\
&&
+ \left[
\frac{9}{2} \epsilon
- \frac{15}{4} \epsilon^2
+ \left[ \frac{1}{8}
- 27 \zeta_3 \right] \epsilon^3
- \left[ \frac{341}{16}
- 99 \zeta_3
+ \frac{81}{2} \zeta_4 \right] \epsilon^4
\right. \nonumber \\
&& \left. ~~~~
- \left[ \frac{1103}{32}
- 42 \zeta_3
+ \frac{297}{2} \zeta_4
- 54 \zeta_5 \right] \epsilon^5
\right. \nonumber \\
&& \left. ~~~~
- \left[ \frac{2161}{64}
- 26 \zeta_3
- 63 \zeta_4
+ 252 \zeta_5
- \frac{135}{2} \zeta_6
- 108 \zeta_3^2 \right] \epsilon^6
\right] \frac{1}{N^2}
\nonumber \\
&& +~ O \left( \epsilon^7, \frac{1}{N^3} \right)
\end{eqnarray}
for the non-singlet case. In comparing both with \cite{25,26} allowance has to
be made for the different conventions in defining $\epsilon$. This is a highly 
non-trivial check for the $O(1/N^2)$ exponents and gives confidence that they 
are correct. Therefore we can derive expressions for the exponents in other 
dimensions. As an example the anomalous dimension of the non-singlet operator 
in two dimensions, for instance, is 
\begin{eqnarray}
\left. \Delta_{\bar{\psi}\sigma^z\psi} \right|_{d=2-2\epsilon} &=&
1 - 2 \epsilon
\nonumber \\
&&
+ \left[
-~ \frac{1}{2}
+ \epsilon
- \zeta_3 \epsilon^3
+ \left[ 2 \zeta_3
- \frac{3}{2} \zeta_4 \right] \epsilon^4
+ \left[ 3 \zeta_4
- 3 \zeta_5 \right] \epsilon^5
+ \left[ 6 \zeta_5
- 5 \zeta_6
- \zeta_3^2 \right] \epsilon^6
\right] \frac{1}{N}
\nonumber \\
&&
+ \left[
-~ \frac{1}{8}
+ \frac{3}{8} \epsilon
+ \left[ -~ \frac{5}{8}
+ \frac{3}{4} \zeta_3 \right] \epsilon^2
+ \left[ \frac{7}{8}
- \frac{9}{2} \zeta_3
+ \frac{9}{8} \zeta_4 \right] \epsilon^3
\right. \nonumber \\
&& \left. ~~~~
+ \left[ \frac{3}{8}
+ \frac{15}{2} \zeta_3
- \frac{27}{4} \zeta_4
+ \frac{3}{2} \zeta_5 \right] \epsilon^4
\right. \nonumber \\
&& \left. ~~~~
+ \left[ -~ \frac{1}{8}
- \frac{7}{2} \zeta_3
+ \frac{45}{4} \zeta_4
- \frac{23}{2} \zeta_5
+ \frac{15}{8} \zeta_6
+ 3 \zeta_3^2 \right] \epsilon^5
\right. \nonumber \\
&& \left. ~~~~
+ \left[ -~ \frac{5}{8}
+ \frac{9}{4} \zeta_7
- \frac{35}{2} \zeta_6
+ \frac{45}{2} \zeta_5
- \frac{21}{4} \zeta_4
+ \frac{5}{2} \zeta_3
+ 9 \zeta_3 \zeta_4
- 17 \zeta_3^2 \right] \epsilon^6
\right] \frac{1}{N^2}
\nonumber \\
&& +~ O \left( \epsilon^7, \frac{1}{N^3} \right) ~.
\end{eqnarray}
One novel feature of the $O(1/N)$ part of this particular exponent is the 
absence of rationals in the coefficients of $\epsilon$ beyond one loop or 
$O(\epsilon)$.

As one of the central motivations for determining the non-singlet exponent 
concerned three dimensions then (\ref{opdim}) implies 
\begin{equation}
\left. \Delta_{\bar{\psi}\sigma^z\psi} \right|_{d=3} ~=~
2 ~-~ \frac{8}{\pi^2 N} ~-~ \frac{16 [ 9 \pi^2 - 100 ]}{9 \pi^4 N^2} ~+~ 
O \left( \frac{1}{N^3} \right)
\end{equation}
or
\begin{equation}
\left. \Delta_{\bar{\psi}\sigma^z\psi} \right|_{d=3} ~=~ 2 ~-~ 
\frac{0.810569}{N} ~+~ \frac{0.203925}{N^2} ~+~ O \left( \frac{1}{N^3} \right)
\label{expqedgnd3}
\end{equation}
numerically. Since the focus on this operator when $N$~$=$~$1$ concerns the 
possible duality connection with the $SU(2)$ symmetric non-compact $CP^1$ sigma
model, setting this value in the three dimensional $O(1/N^2)$ exponent may not 
give a reliable estimate as it is likely to be outside the radius of 
convergence. What would be useful is to use resummation methods in order to see
if the convergence can be improved as well as see to what extent any exponent 
estimate is comparable to those given in \cite{25,26}. While the four 
dimensional renormalization has produced a four loop operator dimension, 
extracting an estimate requires summing the $\epsilon$ expansion of the 
underlying critical exponent and setting $\epsilon$~$=$~$\half$ in our 
$\epsilon$ conventions. Again this choice of $\epsilon$ may be near the radius 
of convergence for extracting an exponent estimate in three dimensions. 
Therefore we have used (\ref{expqedgnd3}) to obtain Pad\'{e} approximants as a 
function of $N$ and evaluated them for various $N$. The numerical estimates are
given in Table $1$. Also included there are Pad\'{e}-Borel estimates which are 
constructed by writing the series (\ref{expqedgnd3}) as a Borel integral and 
then applying a Pad\'{e} approximant to the integrand. The resulting integral
is evaluated numerically and the results displayed in the same Table. In both
the Pad\'{e} and Pad\'{e}-Borel cases there are no singularities either in $N$ 
or the Borel integration parameter. Also included in the final column are the 
four loop perturbative estimates from \cite{26}. What is evident is the close 
agreement of both sets of $O(1/N^2)$ estimates with four loop perturbation 
theory central value down to $N$~$=$~$3$. This is a relatively small value of 
$N$ for which large $N$ results are similar to perturbative estimates compared 
to other exponents in other universality classes. Where there is a clear 
difference is for $N$~$=$~$1$. For the large $N$ case the exponents decrease in
value as $N$ decreases purely due to the negative sign of the $O(1/N)$ term of 
(\ref{opdim}). The perturbative results of \cite{26} do not decrease 
monotonically since the $N$~$=$~$1$ estimate is larger than all the others 
indicated. This may be due to the alternating behaviour of the $\epsilon$ 
series Pad\'{e} approximant with loop order and so a region where the 
approximant converges may not have been reached. The exponent estimate given in
\cite{25} for $N$~$=$~$1$ is $2.12(50)$ which has a central value larger, and 
specifically above $2$, than any of those given in Table $1$. Although the 
Table $1$ values all lie within the error quoted in \cite{25} a value of 
$2.33(1)$ for the same exponent in the dual theory in given in \cite{25}. In 
this case the $N$~$=$~$1$ values of \cite{26} and this article lie well outside
its error. From the large $N$ series (\ref{opdim}) in order to obtain an 
overall value of the exponent above $2$ would require a sizeable positive 
correction from the higher order terms. Overall this would suggest that before 
the duality can be explored more fully then higher order terms, either in large
$N$ or in four dimensional perturbation theory, should be computed. Neither of 
these tasks is a trivial exercise. Alternatively it may be the case that the 
results indicate that there is no duality.

{\begin{table}[ht]
\begin{center}
\begin{tabular}{|c||c||c|c||c|c||c|}
\hline
$N$ & $[0,1]$ P & $[1,1]$ P & $[0,2]$ P & $[1,1]$ PB & $[0,2]$ PB & 
\cite{26} \\
\hline
$1$ & $1.423199$ & $1.352364$ & $1.362789$ & $1.340835$ & $1.418689$ &
$1.98 \pm 0.08$ \\
$2$ & $1.663005$ & $1.640000$ & $1.641745$ & $1.637938$ & $1.656835$ &
$1.74 \pm 0.06$ \\
$3$ & $1.761967$ & $1.750715$ & $1.751288$ & $1.750017$ & $1.757367$ &
$1.76 \pm 0.05$ \\
$4$ & $1.816001$ & $1.809349$ & $1.809603$ & $1.809035$ & $1.812634$ &
$1.81 \pm 0.04$ \\
$5$ & $1.850041$ & $1.845652$ & $1.845787$ & $1.845491$ & $1.847514$ &
$1.84 \pm 0.03$ \\
$6$ & $1.873453$ & $1.870342$ & $1.870421$ & $1.870257$ & $1.871506$ &
$1.86 \pm 0.02$ \\
$10$ & $1.922100$ & $1.920932$ & $1.920950$ & $1.921027$ & $1.921335$ &
$1.917 \pm 0.007$ \\
\hline
\end{tabular}
\end{center}
\begin{center}
{Table $1$. Estimates for 
$\left. \Delta_{\bar{\psi}\sigma^z\psi} \right|_{d=3}$ using Pad\'{e} (P) and
Pad\'{e}-Borel (PB) approximants with comparison to \cite{26}.}
\end{center}
\end{table}}

Finally for the mass operator in the QED-Gross-Neveu universality class in
three dimensions we note that 
\begin{equation}
\left. \Delta_m \right|_{d=3} ~=~
2 ~-~ \frac{8}{\pi^2 N} ~+~ \frac{[ 2464 - 486 \pi^2 ]}{9 \pi^4 N^2} ~+~ 
O \left( \frac{1}{N^3} \right)
\end{equation}
or
\begin{equation}
\left. \Delta_m \right|_{d=3} ~=~ 2 ~-~ \frac{0.810569}{N} ~-~ 
\frac{2.660746}{N^2} ~+~ O \left( \frac{1}{N^3} \right)
\label{massexpqedgnd3}
\end{equation}
numerically. If we follow the prescription indicated in \cite{40} for
extracting an estimate of $N$ for which chiral symmetry breaking occurs, which 
we denote by $\Nc$, we find that $\Nc$~$=$~$3.24$ at leading order but 
$\Nc$~$=$~$4.88$ with the $O(1/N^2)$ correction included. This should be 
compared with the respective values of $4.32$ and $2.85$, \cite{39,40}, for the 
pure QED case. 

\sect{Gross-Neveu universality class.}

In this section we make a brief side step and focus on one of the constituent
theories within (\ref{laguniv}) which is the pure Gross-Neveu model and 
corresponds to omitting terms involving $A_\mu$. While the core critical 
exponents $\eta$ and $\chi$ as well as the fermion mass dimension have been 
already computed to $O(1/N^3)$ and $O(1/N^2)$ respectively in 
\cite{34,43,46,47,48,49,50} that for the non-singlet fermion bilinear operator 
has not been recorded. We do so now as a simple corollary of the formalism of 
the previous section. For completeness we first record the relevant exponents 
and amplitudes to the requisite orders needed to achieve this. While several of
these have been recorded already we repeat them here but with the same spinor 
trace and flavour symmetry group conventions of this article. In previous work,
for instance, two dimensional $\gamma$-matrices were used in the $O(N)$ theory,
\cite{34}. First, the exponents determining the dimensions of the fields are 
\begin{equation}
\eta_1^{\mbox{\footnotesize{GN}}} ~=~ -~ 
\frac{(\mu-1)\Gamma(2\mu-1)}{2\mu\Gamma^3(\mu) \Gamma(1-\mu)}
\end{equation}
and
\begin{equation}
\chi_{\sigma \,1}^{\mbox{\footnotesize{GN}}} ~=~ \frac{\mu}{(\mu-1)}
\eta_1^{\mbox{\footnotesize{GN}}}
\end{equation}
where we append GN to indicate the pure Gross-Neveu universality class.
Consequently the amplitude to $O(1/N^2)$ is determined from 
\begin{eqnarray}
\tilde{z}_1^{\mbox{\footnotesize{GN}}} 
&=& \frac{\Gamma(2\mu-1)} {4 \Gamma^2(\mu) \Gamma(1-\mu)} \nonumber \\
\tilde{z}_2^{\mbox{\footnotesize{GN}}} 
&=& -~ \left[ 
\frac{\mu(2\mu-1)}{2(\mu-1)^2}
\left[ \psi(2 \mu-1) - \psi(1) + \psi(1-\mu) - \psi(\mu-1) \right]
\right. \nonumber \\
&& \left. ~~~~ -~ \frac{1}{(\mu-1)} \right]
\frac{\mu(2\mu-1)\Gamma(\mu)}{2(\mu-1)^2} 
\left( {\eta}_1^{\mbox{\footnotesize{GN}}} \right)^2
\end{eqnarray}
in momentum space which produces 
\begin{eqnarray}
\eta_2^{\mbox{\footnotesize{GN}}} &=& \frac{(2 \mu-1)}{(\mu-1)}
\left[ \psi(2 \mu-1) - \psi(1) + \psi(1-\mu) - \psi(\mu-1) ~-~ 
\frac{1}{2\mu(\mu-1)} 
\right] \left( \eta_1^{\mbox{\footnotesize{GN}}} \right)^2 .~~
\end{eqnarray}

Equipped with these basic building blocks for the large $N$ expansion the
non-singlet bilinear operator dimension is deduced from the computation of
the same quantity in the QED-Gross-Neveu class by formally setting 
$\tilde{y}$~$=$~$0$ at the outset. For example, only the graphs in Figure $5$ 
and $6$ corresponding to $\Gamma_{i\,0n}$ will contribute. At leading order we 
have 
\begin{equation}
\eta_{{\cal O}_{\mbox{ns}}1}^{\mbox{\footnotesize{GN}}} ~=~ \frac{\mu}{(\mu-1)}
\eta_1^{\mbox{\footnotesize{GN}}}
\end{equation}
while the next order produces
\begin{eqnarray}
\eta_{{\cal O}_{\mbox{ns}}2}^{\mbox{\footnotesize{GN}}} &=& 
\frac{\mu (2\mu-1)}{(\mu-1)^2} \left[
\psi(2 \mu-1) - \psi(1) + \psi(1-\mu) - \psi(\mu-1) ~-~ \frac{1}{(\mu-1)} 
\right] \left( {\eta}_1^{\mbox{\footnotesize{GN}}} \right)^2 ~.~~~
\end{eqnarray}
The $\epsilon$ expansion of $\Delta_{\bar{\psi}\sigma^z\psi}$ near four 
dimensions is in full agreement with the corresponding three and four loop 
results of \cite{25,26} when $\bar{g}_2$~$=$~$0$ is set in the operator
anomalous dimension. To assist with comparison for independent computations we
note that 
\begin{eqnarray}
\left. \Delta_{\bar{\psi}\sigma^z\psi}^{\mbox{\footnotesize{GN}}} 
\right|_{d=4-2\epsilon} &=& 3 ~-~ 2 \epsilon
\nonumber \\
&&
+ \left[
\frac{3}{2} \epsilon
- \frac{7}{4} \epsilon^2
- \frac{11}{8} \epsilon^3
+ \left[ 3 \zeta_3 - \frac{19}{16} \right] \epsilon^4
+ \left[ \frac{9}{2} \zeta_4 
- \frac{7}{2} \zeta_3
- \frac{35}{32}
\right] \epsilon^5
\right. \nonumber \\
&& \left. ~~~~
+ \left[ 9 \zeta_5
- \frac{11}{4} \zeta_3
- \frac{21}{4} \zeta_4
- \frac{67}{64} \right] \epsilon^6
\right] \frac{1}{N}
\nonumber \\
&&
+ \left[
-~ \frac{9}{4} \epsilon
+ \frac{147}{16} \epsilon^2
- \frac{71}{32} \epsilon^3
- \left[ \frac{53}{8}
+ 18 \zeta_3
\right] \epsilon^4
- \left[ \frac{65}{8}
- \frac{231}{4} \zeta_3
+ 27 \zeta_4 \right] \epsilon^5
\right. \nonumber \\
&& \left. ~~~~
- \left[ \frac{2127}{256}
+ \frac{37}{8} \zeta_3
- \frac{693}{8} \zeta_4
+ 63 \zeta_5 \right] \epsilon^6
\right] \frac{1}{N^2} ~+~ O \left( \epsilon^7, \frac{1}{N^3} \right)
\end{eqnarray}
near four dimensions while 
\begin{eqnarray}
\left. \Delta_{\bar{\psi}\sigma^z\psi}^{\mbox{\footnotesize{GN}}} 
\right|_{d=2-2\epsilon} &=& 1 ~-~ 2 \epsilon
\nonumber \\
&&
+ \left[
-~ \frac{1}{2} \epsilon
+ \frac{1}{2} \epsilon^2
+ \frac{1}{2} \epsilon^3
+ \left[ \frac{1}{2} 
- \zeta_3 \right] \epsilon^4
+ \left[ \frac{1}{2}
- \frac{3}{2} \zeta_4 
+ \zeta_3 \right] \epsilon^5
\right. \nonumber \\
&& \left. ~~~~
+ \left[ \frac{1}{2} 
- 3 \zeta_5 
+ \frac{3}{2} \zeta_4 
+ \zeta_3 \right] \epsilon^6
\right] \frac{1}{N}
\nonumber \\
&&
+ \left[ 
-~ \frac{1}{4} \epsilon
+ \frac{5}{8} \epsilon^2
+ \frac{1}{8} \epsilon^3
- \left[ \frac{1}{4} 
+ 2 \zeta_3 \right] \epsilon^4
+ \left[ \frac{9}{2} \zeta_3 
- \frac{1}{2} 
- 3 \zeta_4 \right] \epsilon^5
\right. \nonumber \\
&& \left. ~~~~
+ \left[ -~ \frac{5}{8} 
- 7 \zeta_5 
+ \frac{27}{4} \zeta_4 
+ \frac{3}{2} \zeta_3 \right] \epsilon^6
\right] \frac{1}{N^2} ~+~ O \left( \epsilon^7, \frac{1}{N^3} \right)
\end{eqnarray}
in the Gross-Neveu model of \cite{8}. In three dimensions the effect of the
presence or absence of the photon field can be gauged from 
\begin{equation}
\left. \Delta_{\bar{\psi}\sigma^z\psi}^{\mbox{\footnotesize{GN}}} 
\right|_{d=3} ~=~ 2 ~+~ \frac{8}{3\pi^2 N} ~+~ \frac{256}{27\pi^4 N^2} ~+~ 
O \left( \frac{1}{N^3} \right)
\end{equation}
or
\begin{equation}
\left. \Delta_{\bar{\psi}\sigma^z\psi}^{\mbox{\footnotesize{GN}}} 
\right|_{d=3} ~=~ 2 ~+~ \frac{0.270190}{N} ~+~ \frac{0.097337}{N^2} ~+~ 
O \left( \frac{1}{N^3} \right)
\label{expgnd3}
\end{equation}
numerically. Comparing with (\ref{expqedgnd3}) the first order correction is
positive in contrast with (\ref{expgnd3}). This contrasting behaviour has been 
quantified more concretely in Table 2 where we provide Pad\'{e} estimates for 
the same values of $N$ as in Table 1. However we have not recorded estimates
using the Pad\'{e}-Borel method as the positive sign in the leading order
correction in (\ref{expgnd3}) produces singularities in the Pad\'{e}
approximant of the integrand. Although we have no perturbative estimates with 
which to compare it appears there is reasonable agreement of the two $O(1/N^2)$
approximants down to $N$~$=$~$3$. While this is the same value as Table $1$ a 
relatively low value of $N$ for the range of validity is not unreasonable in 
this instance due to the relatively small $O(1/N^2)$ correction. What is 
interesting is that the same value should arise in the QED-Gross-Neveu case as
the analogous correction is roughly three times larger. This is the main
observation of this exercise which was to ascertain for how low a value of $N$ 
one could garner reliable estimates from several orders in $1/N$ for this 
exponent.

{\begin{table}[ht]
\begin{center}
\begin{tabular}{|c||c||c|c||c|c|}
\hline
$N$ & $[0,1]$ & $[1,1]$ & $[0,2]$ \\
\hline
$1$ & $2.312392$ & $2.422339$ & $2.396681$ \\
$2$ & $2.144881$ & $2.164775$ & $2.162517$ \\
$3$ & $2.094310$ & $2.102354$ & $2.101749$ \\
$4$ & $2.069908$ & $2.074233$ & $2.073989$ \\
$5$ & $2.055539$ & $2.058234$ & $2.058112$ \\
$6$ & $2.046069$ & $2.047908$ & $2.047839$ \\
$10$ & $2.027389$ & $2.028029$ & $2.028014$ \\
\hline
\end{tabular}
\end{center}
\begin{center}
{Table $2$. Estimates for 
$\left. \Delta_{\bar{\psi}\sigma^z\psi}^{\mbox{\footnotesize{GN}}} 
\right|_{d=3}$ using Pad\'{e} approximants for the pure Gross-Neveu 
universality class.}
\end{center}
\end{table}}

\sect{Discussion.}

We have completed the $O(1/N^2)$ determination of both flavour non-singlet and 
singlet fermion bilinear operator critical exponents at the Wilson-Fisher fixed
point in the QED-Gross-Neveu universality class. Since the exponents were 
evaluated in $d$-dimensions they bridge between several theories in this class 
including the four dimensional QED-Gross-Neveu-Yukawa theory used in 
\cite{25,26}. As such the $\epsilon$-expansion of both exponents provided 
non-trivial independent checks on the three and four loop of the operator 
anomalous dimension of \cite{25,26}. Once established to be consistent with 
perturbation theory we have provided numerical estimates for the three 
dimensional exponents. This was primarily to inform the debate on the potential
duality connection of the $N$~$=$~$1$ theory with the $SU(2)$-symmetric $CP^1$ 
sigma model where the current focus is on the flavour non-singlet case. As it 
stands both from the perturbative and large $N$ results it would appear that
for that operator dimension more analysis needs to be carried out. From the 
side of the QED-Gross-Neveu universality class one way of improving the 
$N$~$=$~$1$ estimate would be to repeat the approach of \cite{15}. There all 
known data from large $N$ and $\epsilon$-expansions, for instance, were 
combined into improved matched Pad\'{e} approximants up to four loops. This 
provided an interpolating approximation to the critical exponent across the 
dimensions from two to four using the perturbative renormalization group 
information from the critical theories in these dimensions. The form of this 
function for the exponents considered in \cite{15} was not dissimilar to that 
obtained by functional renormalization group methods. Therefore what would be 
useful for the QED-Gross-Neveu universality class is the determination of the 
corresponding renormalization group functions in two dimensions. Although the 
$\beta$-functions are available at two loops, \cite{51,52}, the field and mass 
anomalous dimensions remain to be determined at this and higher loop order. In 
addition to this operator we have provided the fermion mass dimension at
$O(1/N^2)$ which also determines $\chi_\sigma$ as a corollary. This mass
dimension is important in analyses concerning chiral symmetry breaking in the
QED-Gross-Neveu universality class in three dimensions and we have provided an
initial examination of this at $O(1/N^2)$. Further our mass exponent at 
$O(1/N^2)$ should also be useful in supplementing Monte Carlo studies of 
related QED-like theories considered in \cite{53}, for example, where a 
Lagrangian similar to (\ref{lagd4}) is relevant for a confinement transition. 
 
\vspace{1cm}
\noindent
{\bf Acknowledgements.} The author thanks Prof V.P. Gusynin, Dr L. Janssen, 
Prof J. Maciejko, Dr M. Scherer and Prof S. Teber for valuable discussions. All
the authors of \cite{25,26} are also thanked for sharing their results prior to
publication which assisted checks of the exponents. In particular we are 
grateful to Prof J. Maciejko for providing the values of 
$\left. \Delta_{\bar{\psi}\sigma^z\psi} \right|_{d=3}$ for $N$~$\geq$~$3$ 
quoted in Table $1$ which were determined within the analysis of \cite{26} and 
for permission to include them. The graphs were drawn with the {\sc Axodraw}
package \cite{54}. The work was carried out with the support of a DFG Mercator 
Fellowship.

\end{document}